\documentclass{ifacconf}

\usepackage{graphicx}
\graphicspath{{figures/}}

\usepackage{natbib}

\usepackage{amsmath}
\usepackage{amssymb}
\usepackage{mathtools}
\usepackage{physics}

\newcommand{\hodge}{{\star}}

\begin{document}

\begin{frontmatter}
	\title{Exergetic Port-Hamiltonian Systems: Navier-Stokes-Fourier Fluid}

	\author[FAU]{Markus Lohmayer}
	\author[FAU]{Sigrid Leyendecker}

	\address[FAU]{%
		Friedrich-Alexander Universität Erlangen-Nürnberg,
		Erlangen, Germany
		(e-mail: markus.lohmayer@fau.de).
	}

	\begin{abstract} 
		The Exergetic Port-Hamiltonian Systems
modeling language
combines
a graphical syntax inspired by bond graphs
with
a port-Hamiltonian semantics
akin to the GENERIC formalism.
The syntax
enables
the modular and hierarchical specification of
the composition pattern of
lumped and distributed-parameter models.
The semantics
reflects
the first and second law of thermodynamics
as structural properties.
Interconnected and hierarchically defined models
of multiphysical thermodynamic systems
can thus be expressed
in a formal language
accessible to humans and computers alike.
We discuss
a composed model of
the Navier-Stokes-Fourier fluid
on a fixed spatial domain
as an example of an open distributed-parameter system.
At the top level,
the system comprises five subsystems
which model
kinetic energy storage,
internal energy storage,
thermal conduction,
bulk viscosity,
and shear viscosity.

	\end{abstract}

\end{frontmatter}

\section{Introduction}%

Due to the aggravating sustainability crisis,
control- and optimization-oriented modeling
of multiphysical thermodynamic systems
is gaining in importance for
the design of healthy energy systems.
This is but one reason why
port-Hamiltonian models
of open thermodynamic systems,
and in particular of fluid systems,
are of practical interest.

The Navier-Stokes-Fourier fluid
extends
Euler's model of an ideal fluid
with
viscosity and thermal conduction.
%
As a thermodynamic model,
it does not resolve
all relevant details.
Its entropy reflects
incomplete information
about the chaotic motion of atoms
and
entropy growth reflects
irreversible processes
which,
in the absence of external forcing,
drive the system towards equilibrium.
Based on the assumption of local equilibrium,
material properties
pertinent to
internal energy storage%
\footnote{%
	Internal energy is energy stored at
	scales whose details are not resolved by the thermodynamic model.%
}
are defined by a thermodynamic potential
relating
intensive quantities
(e.g.~temperature, pressure)
and
(volumetric densities of) extensive quantities
(e.g.~internal energy, entropy, mass).

\cite{2001SchaftMaschke}
show that
the Euler equations for
an ideal compressible fluid
can be written in port-Hamiltonian form
using a Stokes-Dirac structure.
\cite{%
2021RashadCalifanoSchullerStramigioli1,%
2021RashadCalifanoSchullerStramigioli2}
provide a firm theoretical basis
by deriving this model
via Lie-Poisson reduction,
see Sec.~\ref{ssec:kinetic}.
\cite{2021CalifanoRashadSchullerStramigioli}
extend the model
with
bulk and shear viscosity.
Their Navier-Stokes model
however ignores the thermal domain.
The simplification
hinges on
the barotropicity\footnote{%
	In barotropic flow,
	the mass density is considered to be
	a function of pressure only.%
} assumption.
The goal of the present article is
to continue this line of research
with
a complete thermodynamic model
including also thermal conduction.

Port-Hamiltonian theory
is often considered insufficient
for thermodynamic modeling.
This has motivated related modeling frameworks introduced
by~\cite{2013RamirezMaschkeSbarbaro}
and~\cite{2018SchaftMaschke}.
%
Exergetic Port-Hamiltonian Systems (EPHS)
introduced by~\cite{2021LohmayerKotyczkaLeyendecker}
are port-Hamiltonian systems
which have a thermodynamic structure and interpretation
that is akin to
the GENERIC formalism of nonequilibrium thermodynamics
introduced by~\cite{1997GrmelaOettinger}.
%
The EPHS framework is partly inspired by
the work of~\cite{2018BadlyanMaschkeBeattieMehrmann}
who rewrite
a GENERIC model of the Navier-Stokes-Fourier fluid
in port-Hamiltonian form
using an `exergy-like potential'.
While they rewrite the GENERIC model
into a monolithic port-Hamiltonian system,
our model has explicit compositional structure.
Based on the graphical EPHS syntax,
the exchange of energy
between
model components and boundary ports
is made explicit.
Rather than
using the fluid's internal energy as a state variable,
we choose entropy
since it leads to simpler EPHS models,
see~\cite{2021LohmayerKotyczkaLeyendecker}.
%
%
%
%
Instead of assuming an Euclidian spatial domain,
we work with
a general Riemannian manifold
and
use Cartan's exterior calculus.
Eventually,
this shall enable
a rather general framework
for distributed-parameter port-Hamiltonian systems
and their spatial discretization.

In Sec.~\ref{sec:ephs} and~\ref{sec:exterior}
we provide relevant background on
EPHS and exterior calculus, respectively.
In Sec.~\ref{sec:fluid}
we discuss the fluid model component by component.
In Sec.~\ref{sec:conclusion}
we conclude the article.

\section{Exergetic Port-Hamiltonian Systems}%
\label{sec:ephs}

The EPHS framework
combines port-Hamilton theory with
ideas from the GENERIC formalism and categorical systems theory,
which leads to
a formal modeling language
for multiphysical thermodynamic systems,
see~\cite{2021LohmayerKotyczkaLeyendecker,2022LohmayerLeyendecker}.
%
Using the EPHS syntax,
the composition pattern of
a system and its subsystems
can be expressed in a graphical, yet formal manner.
As syntactic primitive,
a box represents
the typed interface of a system
in terms of its boundary ports.
%
A model semantics
is defined by a functor which assigns
to every box
the set of systems with that interface
and
to every composition pattern
a function which sends
appropriately typed subsystems
to the composite system.

A component for exergy storage
is defined by
a state space $\mathcal{X} \ni x$
and
a Hamiltonian function
$H \colon \mathcal{X} \rightarrow \mathbb{R}$
of the form
\begin{equation}
	\begin{split}
		H(x)
		\: = \:
		E(x) &- E(x_0) \\
		-\theta_0 \bigl( S(x) &- S(x_0) \bigr) \\
		+\pi_0 \bigl( V(x) &- V(x_0) \bigr) \\
		- \mu_0 \bigl( N(x) &- N(x_0) \bigr)
		\,.
	\end{split}
	\label{eq:exergetic_hamiltonian}
\end{equation}%
$E, \, S, \, V, \, N$
are the total energy, entropy, volume and mass.
$\theta_0, \, \pi_0, \, \mu_0$
are the reference temperature, pressure and chemical potential
that define the system's environment.
$H$ is equal to the amount of work $E(x) - E(x_0)$
which may be extracted from the system
in the reversible limit,
implying $S(x) = S(x_0)$,
before the overall system
including its reference environment
reaches the equilibrium state $x_0$.
Hence,
$\theta_0$, $\pi_0$, and $\mu_0$
may be regarded as
Lagrange multipliers
which are constant
because the environment is considered to be infinitely large.
$H$ is a Lyapunov/storage function for the stability of $x_0$.
As the entropy $S$ grows, the exergy $H$ diminishes.
For an isolated EPHS,
$-S$ alone is not a Lyapunov function
because it is unbounded
in the absence of constraints
which state that
$E$, $V$, $N$ are conserved
while $S$ may grow,
see Sec.~3.3 of~\cite{2021LohmayerKotyczkaLeyendecker}
for more details.
Since $H$ is a potential,
we may ignore the constant shifts related to $x_0$.

The Dirac structure of an EPHS
preserves exergetic power,
and thus energy and entropy.
It represents reversible exchange of exergy among components,
akin to ideal wires in electrical circuits.
The resistive structure
is defined such that
energy is conserved
while the net exergetic power
supplied to a dissipative process is non-negative.
Together,
this implies the first and second law of thermodynamics.

\section{Exterior Calculus}%
\label{sec:exterior}

The spatial domain of the fluid is
a smooth, oriented and compact
$n$-dimensional Riemannian manifold $\mathcal{Z}$
with
a possibly non-empty boundary $\partial \mathcal{Z}$
and
inclusion map
$i \colon \partial \mathcal{Z} \xhookrightarrow{} \mathcal{Z}$.
The $C^\infty(\mathcal{Z})$-module
$\Omega^k(\mathcal{Z}) = \Gamma( \wedge^k \, T^* \mathcal{Z})$
of differential $k$-forms
contains
all the anti-symmetric covariant tensors of rank $k$.
$0$-forms are merely smooth functions,
i.e.~$\Omega^0(\mathcal{Z}) = C^\infty(\mathcal{Z})$.
Differential forms
of degrees $0 \leq k \leq n$
form the exterior algebra
whose product
$
\wedge \colon
\Omega^k(\mathcal{Z}) \times \Omega^l(\mathcal{Z})
\rightarrow \Omega^{k+l}(\mathcal{Z})
$
essentially is
the anti-symmetrized tensor product $\otimes$.
A $k$-form can be integrated on
a $k$-dimensional submanifold.
The exterior derivative
$\dd \colon \Omega^k(\mathcal{Z}) \rightarrow \Omega^{k+1}(\mathcal{Z})$
appears in
the Stokes-Cartan theorem
generalizing
the fundamental theorem of calculus ($k=0$),
the classical Stokes theorem ($k=1$),
and the divergence theorem ($k=2$).
It expresses a duality between
differentiation and restriction to the boundary,
i.e.~pullback along the inclusion map $i^*$.
An integration by parts formula follows:
$\forall \, \alpha \in \Omega^k(\mathcal{Z})$
and
$\forall \, \beta  \in \Omega^{(n-k-1)}(\mathcal{Z})$
\begin{equation}
	\begin{split}
		\int_\mathcal{Z} \dd ( \alpha \wedge \beta )
		\: &= \:
		\int_\mathcal{\partial Z} i^* ( \alpha \wedge \beta )
		\: = \:
		\int_\mathcal{\partial Z} i^* \alpha \wedge i^* \beta  \\
		\: &= \:
		\int_\mathcal{Z} \dd \alpha \wedge \beta
		\: + \:
		{(-1)}^k \,
		\int_\mathcal{Z} \alpha \wedge \dd \beta
	\end{split}%
	\label{eq:partial_integration}
\end{equation}
The spatial domain $\mathcal{Z}$
carries a Riemannian metric,
i.e.~a symmetric positive-definite tensor
$g \in \Gamma(T^* \mathcal{X} \otimes T^* \mathcal{X})$
which defines an inner product on every tangent space
and induces further structure:
The musical isomorphism
$\flat \colon T \mathcal{Z} \rightarrow T^* \mathcal{Z}$,
$\sharp \colon T^* \mathcal{Z} \rightarrow T \mathcal{Z}$
identifies
vectors/vector fields and covectors/$1$-forms.
The Hodge star isomorphism
$
\star \colon \Omega^k(\mathcal{Z})
\rightarrow \Omega^{n-k}(\mathcal{Z})
$
establishes a duality between $k$-forms and $(n-k)$-forms.
%
%
Finally,
the covariant derivative $\nabla$
measures the change of a tensor
along a vector.
The Lie derivative $\mathcal{L}$
is metric-independent
and measures the change of a tensor
along the local flow of a vector field.
The interior product
$
\iota_X \colon
\Omega^k(\mathcal{Z})
\rightarrow \Omega^{k-1}(\mathcal{Z})
$
fixes a vector field $X \in \Gamma(T X)$
as the first argument of a $k$-form.
It can be expressed
via the Hodge star,
see~\cite{2003Hirani}:
\begin{equation}
	\begin{split}
		\raisetag{3em}
		\iota_{{(\cdot)}^\sharp} (\cdot) \colon
		\Omega^1(\mathcal{Z}) \times
		\Omega^k(\mathcal{Z})
		&\rightarrow
		\Omega^{k-1}(\mathcal{Z}) \\
		\left( \upsilon, \, \alpha \right)
		&\mapsto
		\iota_{\upsilon^\sharp} \alpha
		= {(-1)}^{(k+1)n} \,
		\hodge (\upsilon \wedge \hodge \alpha)
	\end{split}
	\label{eq:interior_product}
\end{equation}
Cartan's magic formula
$\mathcal{L}_X \alpha = \dd ( \iota_X \alpha ) + \iota_X ( \dd \alpha)$
gives the Lie derivative of a $k$-form $\alpha$
and thereby shows
a duality between exterior derivative and interior product.
We refer to~\cite{1978AbrahamMarsden}
for a relatively detailed introduction
to these concepts.
We define
$\sigma = {(-1)}^{n-1}$.

\section{Fluid Model}%
\label{sec:fluid}

In this section
we present
the fluid model
by discussing
its five subsystems.
Interconnection of
the kinetic energy subsystem
defined in Sec.~\ref{ssec:kinetic}
and
the internal energy subsystem
defined in Sec.~\ref{ssec:internal}
yields
a model of an ideal fluid.
The subsystems defined in Sec.~\ref{ssec:thermal} to~\ref{ssec:shear}
model irreversible processes
and can be added to the overall system independently of each other.

\subsection{Kinetic Energy Subsystem}%
\label{ssec:kinetic}

The kinetic energy subsystem
is central to fluid motion.
A model of
an isolated, ideal, compressible fluid
can be derived
via Lie-Poisson reduction of
Hamiltonian systems on semidirect product Lie groups,
see~\cite{1984MarsdenRatiuWeinstein1,1984MarsdenRatiuWeinstein2}.
The point of departure is
a Lagrangian/material description of fluid motion
formalized as
a curve on the Lie group of automorphisms of $\mathcal{Z}$.
This group
and
a vector space of
advected quantities (e.g.~mass)
together form a semidirect product Lie group
because automorphisms act on advected quantities.
The reduction procedure takes
the Lagrangian description
on the semidirect product Lie group
to an Eulerian/spatial description
on the dual of its semidirect product Lie algebra.
A Lie algebra isomorphism
between the reduced phase space
and a space of differential forms
enables
the straightforward generalization
of the Lie-Poisson structure
to a Stokes-Dirac structure
and thus also the generalization of
the Hamiltonian model
of ideal compressible fluid motion
to a port-Hamiltonian model,
see~\cite{2021RashadCalifanoSchullerStramigioli1}.
This permits
the consideration of spatial domains with permeable boundaries
and
it enables
the extension of the model
through interconnection with other port-Hamiltonian systems.

The composition pattern of the subsystem
is given by
the following expression
in the EPHS syntax:
\begin{center}
	\includegraphics[height=4.3cm]{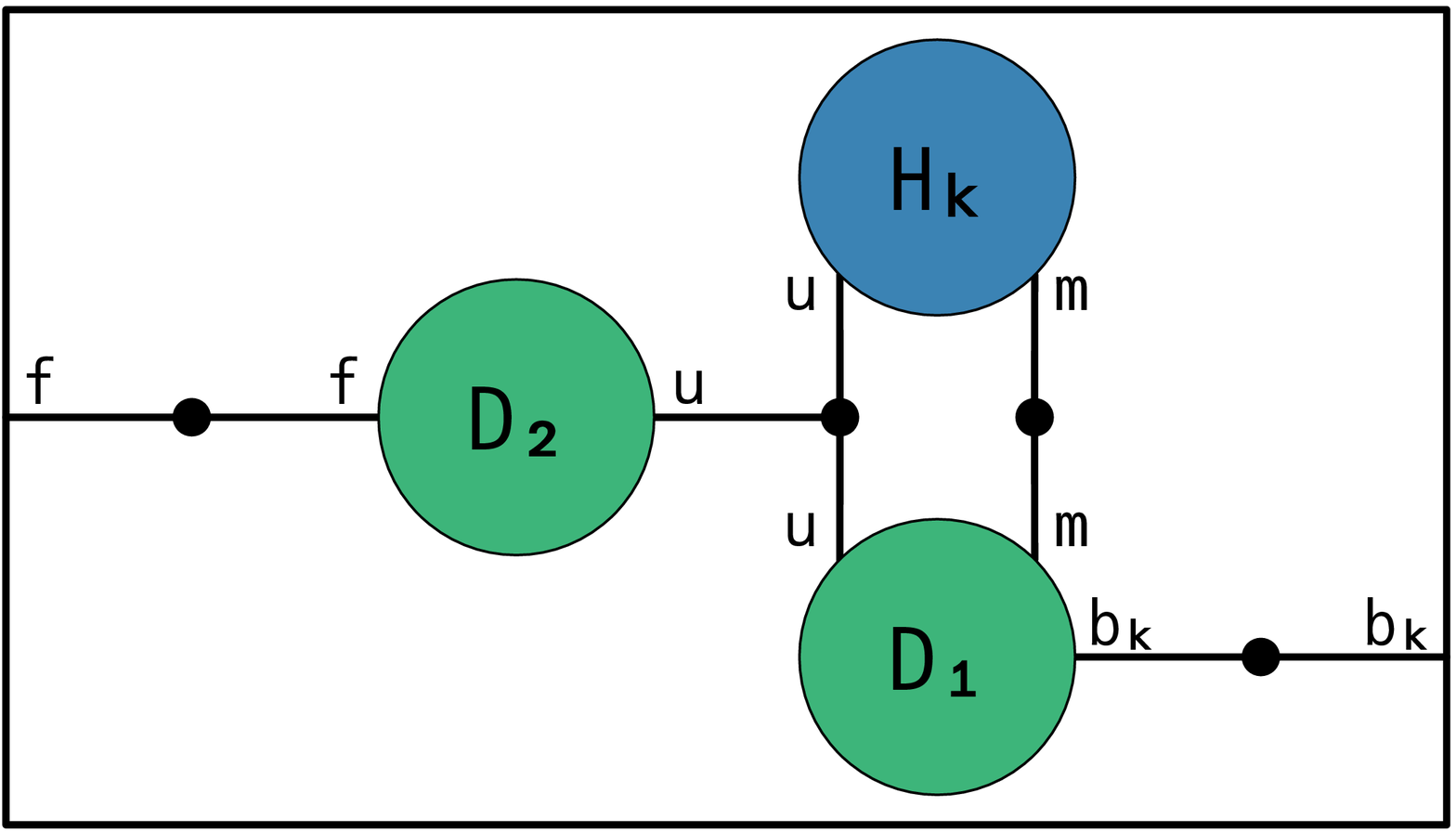}
\end{center}


First, we discuss the semantic data
associated to
box $\texttt{H}_\texttt{k}$
representing storage of kinetic energy.
Linear momentum
and advected mass
are the canonical phase space variables.
The momentum $p$ is a $1$-form
since its change is equal to
the force $1$-form
that is (energetically) dual to the velocity vector field.
The integral over $\mathcal{Z}$
of the mass $n$-form $m$
yields the total fluid mass.
Rather than momentum,
the model uses
the velocity (or specific momentum)
$1$-form $\upsilon = p / \hodge m$
as a state variable,
since it leads to
simpler equations
and presumably has
advantages for numerical computation.
Thus, the energy variables are
\begin{equation}
	\left( \upsilon, \, m \right)
	\in
	\mathcal{X} =
	\Omega^1(\mathcal{Z}) \times
	\Omega^n(\mathcal{Z})
	\,.
	\label{eq:kinetic_state}
\end{equation}
The Hamiltonian
$\mathcal{X} \rightarrow \mathbb{R}$
yields the total kinetic energy:
\begin{equation}
	\begin{split}
		H(\upsilon, \, m)
		\: &= \:
		\int_\mathcal{Z} \frac{1}{2} \, m \, g(\upsilon^\sharp, \, \upsilon^\sharp)
		\: = \:
		\int_\mathcal{Z} \frac{1}{2} \, m \, \iota_{\upsilon^\sharp} \upsilon
		\\
		\: &\overset{\eqref{eq:interior_product}}{=} \:
		\int_\mathcal{Z} \frac{1}{2} \, m \,
			\hodge (\upsilon \wedge \hodge \upsilon)
		\: = \:
		\int_\mathcal{Z} \frac{1}{2} \, \hodge m \,
			\upsilon \wedge \hodge \upsilon
	\end{split}%
	\label{eq:kinetic_hamiltonian}
\end{equation}
The partial variational derivative
$\delta_\upsilon H$
e.g.~%
is defined by
$
H(\upsilon + \epsilon \, \delta \upsilon, \, m) =
H(\upsilon, \, m) +
\epsilon \, \int_\mathcal{Z} \delta_\upsilon H \wedge \delta \upsilon +
\mathcal{O}(\epsilon^2)
$
for $\epsilon \rightarrow 0$.
We obtain the following partial derivatives:
%
\begin{subequations}
	\begin{align}
		\delta_\upsilon H
		\: &= \:
		\sigma \, \hodge m \, \hodge \upsilon
		\: \overset{\eqref{eq:interior_product}}{=} \:
		\sigma \, \iota_{\upsilon^\sharp} m
		\: \in \: \Omega^{n-1}(\mathcal{Z})%
		\label{eq:kinetic_storage_effort_upsilon} \\
		\delta_m H
		\: &= \:
		\frac{1}{2} \, \hodge (\upsilon \wedge \hodge \upsilon)
		\: \overset{\eqref{eq:interior_product}}{=} \:
		\frac{1}{2} \, \iota_{\upsilon^\sharp} \upsilon
		\: \in \: \Omega^0(\mathcal{Z})%
		\label{eq:kinetic_storage_effort_m}
	\end{align}%
	\label{eq:kinetic_storage_effort}%
\end{subequations}
The box $\texttt{H}_\texttt{k}$ has ports
\texttt{u}  and \texttt{m}
related to
$\upsilon$ and $m$, respectively.
Each port has
a flow variable \texttt{f}
and an effort variable \texttt{e}.
The port variables are defined by
\begin{equation}
	\left(
		\texttt{u.f}, \,
		\texttt{m.f}, \,
		\texttt{u.e}, \,
		\texttt{m.e}
	\right)
	\: = \:
	\left(
		\dot{\upsilon}, \, \dot{m}, \,
		\delta_\upsilon H, \, \delta_m H
	\right).
	\label{eq:kinetic_storage_port}
\end{equation}


We now discuss
box $\texttt{D}_\texttt{1}$
which represents
the Stokes-Dirac structure
generalizing the original Lie-Poisson structure.
Ports
\texttt{u} and \texttt{m}
are related to
$\upsilon$ and $m$,
while
$\texttt{b}_\texttt{k}$ is related to
energy exchange across $\partial \mathcal{Z}$.
The port variables are constrained
by the following relation:
\begin{subequations}
	\begin{align}
		\left[
			\begin{array}{c}
				\texttt{u.f} \\
				\texttt{m.f} \\
			\end{array}
		\right]
		\: &= \:
		\left[
			\begin{array}{cc}
				S(\cdot) & \dd(\cdot) \\
				\sigma \, \dd(\cdot) & 0 \\
			\end{array}
		\right]
		\,
		\left[
			\begin{array}{c}
				\texttt{u.e} \\
				\texttt{m.e} \\
			\end{array}
		\right]%
		\label{eq:kinetic_dirac_domain}
		\\
		\left[
			\begin{array}{c}
				\texttt{b}_\texttt{k} \texttt{.f} \\
				\texttt{b}_\texttt{k} \texttt{.e}
			\end{array}
		\right]
		\: &= \:
		\left[
			\begin{array}{cc}
				-\sigma \, i^*(\cdot) & 0 \\
				 0 & i^*(\cdot)
			\end{array}
		\right]
		\,
		\left[
			\begin{array}{c}
				\texttt{u.e} \\
				\texttt{m.e}
			\end{array}
		\right]%
		\label{eq:kinetic_dirac_boundary}
	\end{align}%
	%
	\begin{equation}
		\text{where }
		S(\cdot)
		\: = \:
		\frac{1}{\hodge m} \,
		\iota_{(\sharp \circ \hodge)(\cdot)}\dd \upsilon
		\: \overset{\eqref{eq:interior_product}}{=} \:
		-\sigma \, \frac{1}{\hodge m} \,
		\hodge \bigl( \hodge (\cdot) \wedge \hodge \dd \upsilon \bigr)
		\label{eq:kinetic_dirac_self_advection}
	\end{equation}
	\label{eq:kinetic_dirac}
\end{subequations}
The Dirac structure implies
a power-balance equation:
\begin{equation*}
	\int_\mathcal{Z} \bigl(
		\texttt{u.e} \wedge \texttt{u.f}
		+
		\texttt{m.e} \wedge \texttt{m.f}
	\bigr)
	\: + \:
	\int_{\partial \mathcal{Z}}
	\texttt{b}_\texttt{k} \texttt{.e} \wedge
	\texttt{b}_\texttt{k} \texttt{.f}
	\: = \: 0
\end{equation*}

Box $\texttt{D}_\texttt{2}$
represents a Dirac structure
which transforms
the force acting on the fluid via port \texttt{f}
to a change in velocity:
\begin{equation}
	\left[
		\begin{array}{c}
			\texttt{u.f} \\
			\hline
			\texttt{f.e}
		\end{array}
	\right]
	\: = \:
	\left[
		\begin{array}{c|c}
			0 & -\sigma / \hodge m  \\
			\hline
			\sigma / \hodge m & 0
		\end{array}
	\right]
	\,
	\left[
		\begin{array}{c}
			\texttt{u.e} \\
			\hline
			\texttt{f.f}
		\end{array}
	\right]%
	\label{eq:kinetic_dirac_force}
\end{equation}


The composition pattern
states
the interconnection of
components
$\texttt{H}_\texttt{k}$,
$\texttt{D}_\texttt{1}$,
$\texttt{D}_\texttt{2}$
and boundary ports
\texttt{f}, $\texttt{b}_\texttt{k}$.
Here, `boundary' refers to
the system boundary
represented by
the outer box of the expression.
This is orthogonal to
the concept of
distributed versus boundary ports
in the sense of $\mathcal{Z}$ vs $\partial \mathcal{Z}$
since
different systems
may be defined on the same spatial domain
and
the definition of a single system
may involve different spatial domains.
%
Ports are connected to
$0$-junctions
where flow variables sum to zero and effort variables are equal.
E.g.~%
$
\texttt{H}_\texttt{k}\texttt{.u.f} +
\texttt{D}_\texttt{1}\texttt{.u.f} +
\texttt{D}_\texttt{2}\texttt{.u.f} = 0
$
and
$
\texttt{H}_\texttt{k}\texttt{.u.e} =
\texttt{D}_\texttt{1}\texttt{.u.e} =
\texttt{D}_\texttt{2}\texttt{.u.e}
$.
Because naming is hard,
we use the same labels
for ports of different boxes
connected to the same junction.
This shall also apply to the interconnection of the five subsystems.
By applying the composition pattern to
the semantic data associated to the three inner boxes,
we obtain the semantic data associated to the composite subsystem
filling the outer box:
\begin{subequations}
	\begin{align}
		\dot{\upsilon}
		\: &= \:
		-\iota_{\upsilon^\sharp}(\dd \upsilon)
		-\dd (\frac{1}{2} \hodge (\upsilon \wedge \hodge \upsilon))
		+\frac{\sigma}{\hodge m} \, \texttt{f.f}%
		\label{eq:kinetic_momentum_balance}
		\\
		\dot{m}
		\: &= \:
		-\dd (\hodge m \, \hodge \upsilon)
		\: = \:
		-\mathcal{L}_{v^\sharp} m%
		\label{eq:kinetic_mass_balance}
		\\
		\texttt{f.e}
		\: &= \:
		\hodge \upsilon%
		\label{eq:kinetic_output}
		\\
		\texttt{b}_\texttt{k} \texttt{.f}
		\: &= \:
		-i^*(\hodge m \, \hodge \upsilon)
		\label{eq:kinetic_boundary_f}
		\\
		\texttt{b}_\texttt{k} \texttt{.e}
		\: &= \:
		i^* \bigl( \hodge (\upsilon \wedge \hodge \upsilon) / 2 \bigr)%
		\label{eq:kinetic_boundary_e}
	\end{align}%
	\label{eq:kinetic_resulting_eqs}%
\end{subequations}
The Lie derivative appears in~\eqref{eq:kinetic_mass_balance}
since $m$ is advected.
While
\texttt{f.f} is the force acting on a fluid particle,
\texttt{f.e} is the velocity (or volume flux).
While
$\texttt{b}_\texttt{k}\texttt{.f}$ is the mass influx,
$\texttt{b}_\texttt{k}\texttt{.e}$ is the kinetic energy per unit of mass.

\subsection{Internal Energy Subsystem}%
\label{ssec:internal}

While
kinetic energy comes from macroscopic motion,
internal energy corresponds to microscopic motion,
which at the macroscopic scale manifests as
temperature and pressure.
The composition pattern
of the internal energy subsystem
is given by the following expression:
\begin{center}
	\includegraphics[height=4.3cm]{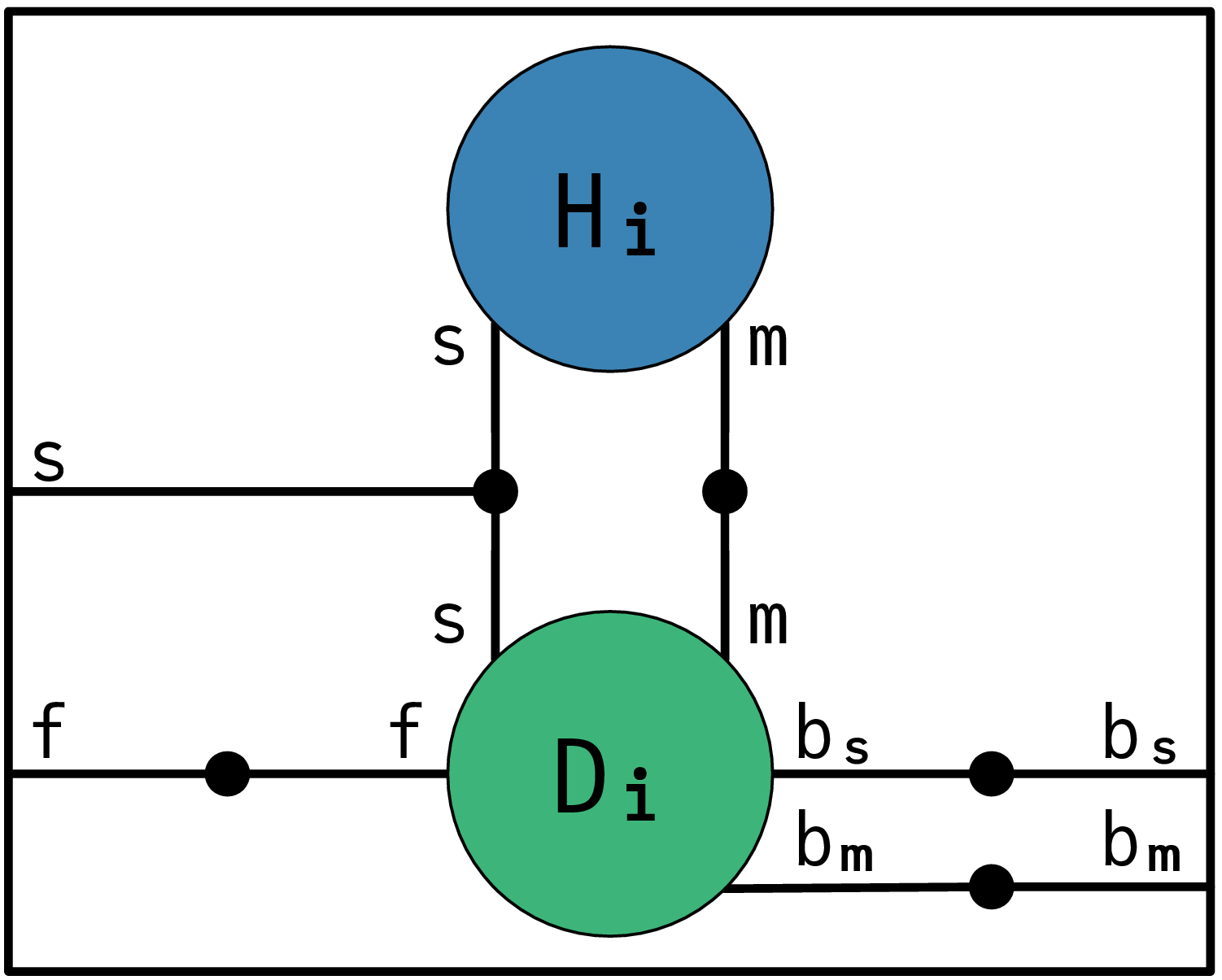}
\end{center}


We first discuss the semantic data associated to
box $\texttt{H}_\texttt{i}$ representing
storage of internal energy.
The entropy $s$
reflects incomplete information
about the state of the fluid
at more microscopic scales
which are not resolved by the model.
Assuming local equilibrium,
thermodynamic properties of the fluid
are defined by a potential $U$
which yields the internal energy
as a function of
entropy density and mass density.
The component's energy variables are
\begin{equation}
	\left( s, \, m \right)
	\in
	\mathcal{X} =
	\Omega^n(\mathcal{Z}) \times
	\Omega^n(\mathcal{Z})
	\,.
	\label{eq:internal_state}
\end{equation}
The Hamiltonian
$H \colon \mathcal{X} \rightarrow \mathbb{R}$
yields, modulo a constant, the total exergy of the fluid
due to its internal energy:
\begin{equation}
	H(s, \, m)
	\: = \:
	\int_\mathcal{Z} \bigl(
		\hodge U(\hodge s, \, \hodge m) - \theta_0 \, s - \mu_0 \, m
	\bigr)
	\label{eq:internal_hamiltonian}
\end{equation}
The intensive variables
temperature $\theta$,
chemical potential $\mu$
and pressure $\pi$
take values in $\Omega^0(\mathcal{Z})$
and are defined by
\begin{subequations}
	\begin{align}
		\theta
		\: &= \:
		\pdv{U(\hodge s, \, \hodge m)}{(\hodge s)}
		\label{eq:theta} \\
		\mu
		\: &= \:
		\pdv{U(\hodge s, \, \hodge m)}{(\hodge m)}
		\label{eq:mu} \\
		\pi
		\: &= \:
		\theta \, \hodge s + \mu \, \hodge m - U(\hodge s, \, \hodge m)
		\,.
		\label{eq:pi}
	\end{align}%
	\label{eq:intensive}%
\end{subequations}
Eq.~\eqref{eq:pi} follows from
$1$-homogeneity of $U$.
The partial variational derivatives are consequently given by
\begin{subequations}
	\begin{align}
		\delta_s H
		\: &= \:
		\theta - \theta_0
		\: \in \: \Omega^0(\mathcal{Z})%
		\label{eq:internal_storage_effort_s} \\
		\delta_m H
		\: &= \:
		\mu - \mu_0
		\: \in \: \Omega^0(\mathcal{Z})%
		\label{eq:internal_storage_effort_m}
	\end{align}%
	\label{eq:internal_storage_effort}
\end{subequations}
and the port variables of $\texttt{H}_\texttt{i}$ are defined by
\begin{equation}
	\left(
		\texttt{s.f}, \,
		\texttt{m.f}, \,
		\texttt{s.e}, \,
		\texttt{m.e}
	\right)
	\: = \:
	\left(
		\dot{s}, \, \dot{m}, \,
		\delta_s H, \, \delta_m H
	\right).
	\label{eq:internal_storage_port}
\end{equation}


Box $\texttt{D}_\texttt{i}$ represents
the reversible coupling of
fluid motion
and
internal energy storage.
Since the boundary ports labeled \texttt{f}
of the kinetic and internal energy subsystems
are connected,
\texttt{f.e}
is the fluid velocity
that advects $s$, $m$
and hence also the internal energy,
while \texttt{f.f}
is the force
which results from the internal energy
and affects fluid motion.
Ports
$\texttt{b}_\texttt{m}$
and
$\texttt{b}_\texttt{s}$
are related to
advection of internal energy
across $\partial \mathcal{Z}$.
The port variables of $\texttt{D}_\texttt{i}$
are constrained by
the following relation
defining a Stokes-Dirac structure:
%
\begin{subequations}
	\begin{align}
		\left[
			\begin{array}{c}
				\texttt{s.f} \\
				\texttt{m.f} \\
				\texttt{f.f}
			\end{array}
		\right]
		&=
		\left[
			\begin{array}{ccc}
				0 & 0 & \dd ( \hodge s \; \cdot) \\
				0 & 0 & \dd ( \hodge m \; \cdot) \\
				\sigma \, \hodge s \dd(\cdot) & \sigma \, \hodge m \dd(\cdot) & 0
			\end{array}
		\right]
		\left[
			\begin{array}{c}
				\texttt{s.e} \\
				\texttt{m.e} \\
				\texttt{f.e}
			\end{array}
		\right]%
		\label{eq:internal_dirac_domain}
		\\
		\left[
			\begin{array}{c}
				\texttt{b}_\texttt{s} \texttt{.f} \\
				\texttt{b}_\texttt{s} \texttt{.e} \\
				\texttt{b}_\texttt{m} \texttt{.f} \\
				\texttt{b}_\texttt{m} \texttt{.e}
			\end{array}
		\right]
		&=
		\left[
			\begin{array}{ccc}
				0 & 0 & -i^*( \hodge s \; \cdot) \\
				i^*(\cdot)  & 0 & 0 \\
				0 & 0 & -i^*( \hodge m \; \cdot) \\
				0 & i^*(\cdot)  & 0
			\end{array}
		\right]
		\,
		\left[
			\begin{array}{c}
				\texttt{s.e} \\
				\texttt{m.e} \\
				\texttt{f.e}
			\end{array}
		\right]%
		\label{eq:internal_dirac_boundary}
	\end{align}%
	\label{eq:internal_dirac}
\end{subequations}


The composition pattern states
the interconnection of
components
$\texttt{H}_\texttt{i}$,
$\texttt{D}_\texttt{i}$,
and boundary ports
\texttt{f},
\texttt{s},
$\texttt{b}_\texttt{s}$,
$\texttt{b}_\texttt{m}$.
Its application to
the semantic data associated to
$\texttt{H}_\texttt{i}$ and
$\texttt{D}_\texttt{i}$
yields
semantic data associated to the entire subsystem:
\begin{subequations}
	\begin{align}
		\dot{s}
		\: &= \:
		-\dd ( \hodge s \; \texttt{f.e} ) + \texttt{s.f}
		\label{eq:internal_entropy_balance}
		\\
		\dot{m}
		\: &= \:
		-\dd ( \hodge m \; \texttt{f.e} )
		\label{eq:internal_mass_balance}
		\\
		\texttt{f.f}
		\: &= \:
		\sigma \, \dd \pi%
		\label{eq:internal_force_output}%
		\\
		\texttt{s.e}
		\: &= \:
		\theta - \theta_0%
		\label{eq:internal_temp_output}%
		\\
		\texttt{b}_\texttt{s} \texttt{.f}
		\: &= \:
		-i^* ( \hodge s \, \texttt{f.e} )%
		\label{eq:internal_boundary_s_f}
		\\
		\texttt{b}_\texttt{s} \texttt{.e}
		\: &= \:
		i^*(\theta - \theta_0)
		\label{eq:internal_boundary_s_e}
		\\
		\texttt{b}_\texttt{m} \texttt{.f}
		\: &= \:
		-i^* ( \hodge m \, \texttt{f.e} )%
		\label{eq:internal_boundary_m_f}
		\\
		\texttt{b}_\texttt{m} \texttt{.e}
		\: &= \:
		i^*(\mu - \mu_0)
		\label{eq:internal_boundary_m_e}
	\end{align}%
	\label{eq:internal_resulting_eqs}%
\end{subequations}
Terms with $\texttt{f.e} = \hodge \upsilon$
in~\eqref{eq:internal_entropy_balance}
and~\eqref{eq:internal_mass_balance}
are Lie derivatives
since $s$, $m$ are advected quantities.
Eq.~\eqref{eq:internal_force_output} follows from
\begin{equation}
		\dd \pi
		\: = \:
		\hodge s \, \dd \theta +
		\hodge m \, \dd \mu
	\label{eq:differential_pi}
\end{equation}
which in turn is a consequence of~\eqref{eq:intensive}.
While
\texttt{f.e} is the fluid velocity,
\texttt{f.f} is the force resulting from thermodynamic pressure.
Port \texttt{s} is relevant for adding irreversible processes.
While
\texttt{s.f} is the local change in entropy caused by irreversible processes,
\texttt{s.e} is the relative temperature.
While
$\texttt{b}_\texttt{s}\texttt{.f}$ is the entropy influx,
$\texttt{b}_\texttt{s}\texttt{.e}$ is the relative temperature at the boundary.
Similarly,
$\texttt{b}_\texttt{m}\texttt{.f}$ is the mass influx
and
$\texttt{b}_\texttt{m}\texttt{.e}$ is the relative chemical potential at the boundary.

\subsection{Thermal Conduction}%
\label{ssec:thermal}

%
Thermal conduction
is a relaxation process
counteracting
nonuniform temperature
with a diffusive heat flux.
The composition pattern of
the thermal conduction subsystem
is given by
the following identity expression:
\begin{center}
	\includegraphics[width=5.3cm]{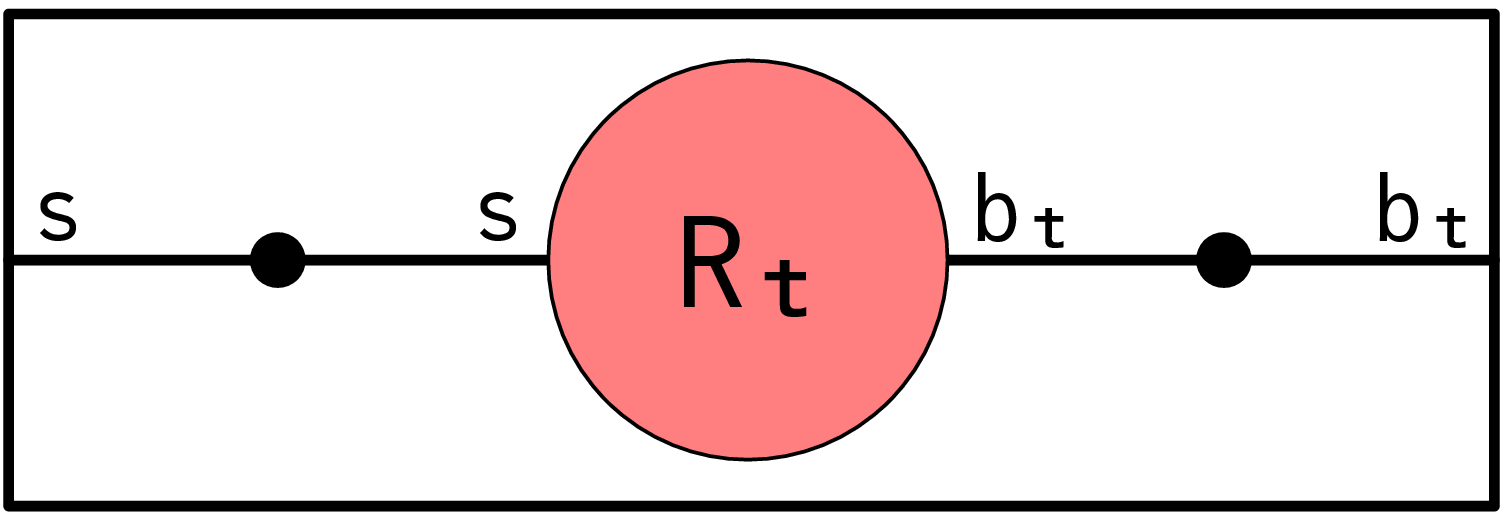}
\end{center}
Ports \texttt{s}
and $\texttt{b}_\texttt{t}$
are related to
entropy flux
due to thermal conduction
within $\mathcal{Z}$ and across $\partial \mathcal{Z}$,
respectively.
Box $\texttt{R}_\texttt{t}$
represents the irreversible process.
Its port variables
are constrained by
the following relation
defining a
resistive structure:
%
\begin{subequations}
	\begin{align}
		\left[
			\texttt{s.f}
		\right]
		\: &= \:
		\left[
			-\frac{1}{\theta} \,
			\dd (
				\hodge \biggl(
					\kappa \,
					\frac{1}{\theta_0} \,
					\theta^2 \,
					\dd (\frac{1}{\theta} \, \cdot )
				\biggr)
			)
		\right]
		\,
		\left[
			\texttt{s.e}
		\right]
		\label{eq:thermal_domain}
		\\
		\left[
			\begin{array}{c}
				\texttt{b}_\texttt{t} \texttt{.f} \\
				\texttt{b}_\texttt{t} \texttt{.e}
			\end{array}
		\right]
		\: &= \:
		\left[
			\begin{array}{cc}
				0 & i^*\biggl(
				\frac{1}{\theta} \,
				\hodge \biggl(
					\kappa \,
					\frac{1}{\theta_0} \, \theta^2 \,
					\dd (\frac{1}{\theta} \, \cdot )
				\biggr)
				\biggr)
				\\
				i^*(\cdot) & 0
			\end{array}
		\right]
		\,
		\left[
			\texttt{s.e}
		\right]%
		\label{eq:thermal_boundary}
		\raisetag{1.1em}
	\end{align}%
	\label{eq:thermal}%
\end{subequations}
%
To make sense of~\eqref{eq:thermal},
we briefly forget about fluid motion
and connect
port $\texttt{R}_\texttt{t}\texttt{.s}$
with
a component $\texttt{H}_\texttt{i}$
representing
the thermal capacity of
a rigid material.
$\texttt{H}_\texttt{i}$ is defined by
a Hamiltonian of the form
$H(s) = \int_\mathcal{Z} \left( \hodge U(\hodge s) - \theta_0 \, s \right)$
and thus we have
$\delta_s H = \theta - \theta_0$.
The first law then says that
the change of internal energy
$u \in \Omega^n(\mathcal{Z})$
in $\mathcal{Z}$
is due to
the heat flux
$\phi \in \Omega^{n-1}(\mathcal{Z})$
across
the boundary
$\partial \mathcal{Z}$, i.e.
\begin{equation*}
	\int_\mathcal{Z} \dot{u}
	\: = \:
	\int_{\partial \mathcal{Z}} i^*(-\phi)
	\: = \:
	\int_\mathcal{Z} \dd (-\phi)
\end{equation*}
The minus sign
comes from
the induced Stokes orientation on $\partial \mathcal{Z}$
which makes $\phi$
the outgoing heat flux.
We hence have the local energy balance
$\dot{u} = \dd(-\phi)$.
Fourier's law gives
$\phi = -\kappa \, \hodge \, \dd \theta$.
The thermal conductivity $\kappa \geq 0$
is constant
or depends on $\theta$.
Using
$
\dot{u} =
\pdv{(\hodge U(\hodge s))}{(\hodge s)} \wedge \dot{s} =
\theta \, \dot{s}
$,
we obtain
$
\dot{s} =
\frac{1}{\theta} \, \dot{u} =
\frac{1}{\theta} \, \dd(-\phi) =
\frac{1}{\theta} \, \dd( \kappa \, \hodge \, \dd \theta )
$.
We get the same local entropy balance
from~\eqref{eq:thermal_domain}
and the interconnection of
$\texttt{H}_\texttt{i}\texttt{.s}$ and
$\texttt{R}_\texttt{t}\texttt{.s}$
yielding
$\dot{s} + \texttt{s.f} = 0$
and
$\theta - \theta_0 = \texttt{s.e}$.
To see this,
we identify
the heat flux
in~\eqref{eq:thermal_domain}:
\begin{equation*}
	\begin{split}
		-\phi
		\: &= \:
		\hodge \biggl(
			\kappa \,
			\frac{1}{\theta_0} \, \theta^2 \,
			\dd (\frac{1}{\theta} \, (\theta - \theta_0) )
		\biggr)
		\\
		\: &= \:
		-\hodge \biggl(
			\kappa \,
			\theta^2 \,
			\dd (\frac{1}{\theta} )
		\biggr)
		\: = \:
		\hodge \bigl(
			\kappa \,
			\dd \theta
		\bigr)
		\: = \:
		\kappa \,
		\hodge \dd \theta
	\end{split}
\end{equation*}
%
To see that
energy is conserved at the resistive structure,
we note that
$\texttt{s.e} = \theta - \theta_0$
where
$\theta$ stems from the energy $E$
and
the shift
$-\theta_0$ stems from the entropy $S$,
see
the first and second line
of~\eqref{eq:exergetic_hamiltonian}.
By omitting the shift,
i.e.~by letting $\texttt{s.e} = \theta$,
we assert that
$E$ is a metric Casimir,
meaning that
the irreversible process conserves energy.
%
The exergy balance of
the system consisting of
$\texttt{H}_\texttt{i}$ and
$\texttt{R}_\texttt{t}$
reads
\begin{equation}
	\begin{split}
		\raisetag{4.5em}
		\dot{H} =
		&\int_\mathcal{Z} \delta_s H \wedge \dot{s} =
		\int_\mathcal{Z} \texttt{s.e} \wedge (-\texttt{s.f}) =
		\int_\mathcal{Z} \frac{\theta - \theta_0}{\theta} \, \dd (-\phi)
		\\
		= &\int_{\partial \mathcal{Z}}
		i^*(\theta - \theta_0) \wedge i^* \biggl(-\frac{1}{\theta} \, \phi \biggr)
		\\
		&-\theta_0 \, \int_\mathcal{Z}
		\frac{1}{\theta_0} \dd (
			\frac{1}{\theta} \,
			\texttt{s.e}
		)
		\wedge
		\hodge \biggl(
			\kappa \,
			\theta^2 \,
			\frac{1}{\theta_0}
			\dd (\frac{1}{\theta} \, \texttt{s.e} )
		\biggr)
		\,.
	\end{split}
	\label{eq:thermal_exergy_balance}
\end{equation}
Eq.~\eqref{eq:thermal_exergy_balance} says that
the stored exergetic power is equal to
the supplied exergetic power
minus the lost exergetic power.
In accordance with the second law,
the last line shows that
the lost exergetic power
is non-negative.
The first line shows that
the stored exergetic power
is equal to the
Carnot efficiency multiplied by
the net incoming heat flux.
While the first and second line
are related to the strong form
of the evolution equation
together with boundary conditions,
the last line is related to the weak form
which implicitly includes boundary conditions.
The weak form shows that
the resistive structure is
defined by a symmetric operator
in accordance with
Onsager's reciprocal relations.
%
By analyzing the structure of this symmetric operator
we can identify
the thermodynamic force and flux.
The term
$
\frac{1}{\theta_0} \dd (\frac{1}{\theta} \, \texttt{s.e} )
= -\dd(\frac{1}{\theta})
$
is called the thermodynamic force
since it drives the relaxation process.
It is obtained from
the differential (i.e.~variational derivative)
of the exergetic Hamiltonian
given by
\texttt{s.e}
via the operator
$\frac{1}{\theta_0} \dd (\frac{1}{\theta} \, \cdot )$
which by construction
cancels the energetic contribution.
This means that entropy
is driving the relaxation dynamics
and energy is left invariant.
The operator
$\hodge( \kappa \, \theta^2 \, \cdot)$
turns
the thermodynamic force into
the thermodynamic flux
which in this case is the heat flux $-\phi$.
The integrand is
the product of
force and flux,
which is equal to
the entropy production rate.
Multiplication with the leading factor $\theta_0$
finally yields the exergy destruction rate.

\subsection{Bulk Viscosity}%
\label{ssec:bulk}

%
Bulk viscosity is a relaxation process
counteracting
local changes in volume
and
thus
in particular leads to damping of acoustic waves.
The composition pattern of
the bulk viscosity subsystem
is given by the following expression:
\begin{center}
	\includegraphics[width=5.3cm]{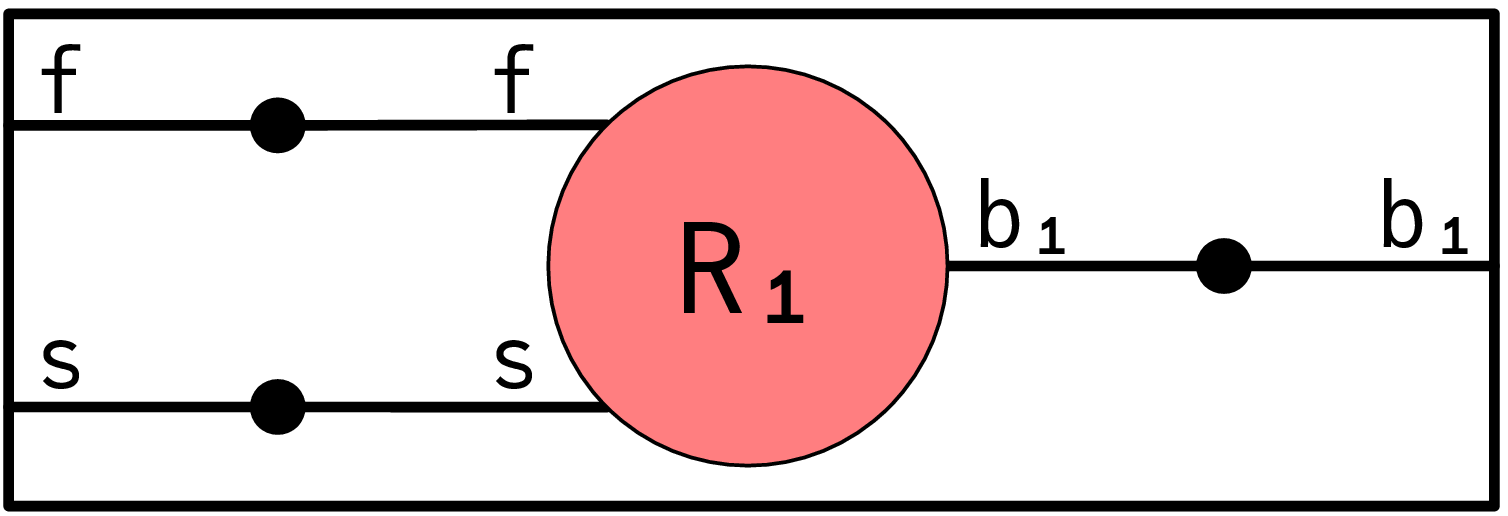}
\end{center}
Port \texttt{f}
is related to the viscous force
counteracting volume change,
port \texttt{s}
is related to the corresponding increase in entropy,
and port $\texttt{b}_\texttt{1}$
is related to the viscous force acting
across $\partial \mathcal{Z}$.
The port variables of $\texttt{R}_\texttt{1}$
are constrained by the following relation
defining a resistive structure:
\begin{subequations}
	\begin{align}
		\raisetag{2em}
		\left[
			\begin{array}{c}
				\texttt{f.f} \\
				\texttt{s.f}
			\end{array}
		\right]
		&=
		\frac{1}{\theta_0}
		\left[
			\begin{array}{cc}
				-\sigma \dd \bigl( \mu_b \, \hodge \dd(\cdot) \theta \bigr) &
				\sigma \dd \bigl( \mu_b (\hodge \dd \hodge \upsilon) \, (\cdot) \bigr) \\
				-\mu_b \, \hodge \dd(\cdot) \, \dd \hodge \upsilon &
				\frac{1}{\theta} \mu_b (\hodge \dd \hodge \upsilon) \, \dd \hodge \upsilon \, (\cdot)
			\end{array}
		\right]
		\left[
			\begin{array}{c}
				\texttt{f.e} \\
				\texttt{s.e}
			\end{array}
		\right]
		\label{eq:bulk_domain}
		\\
		\begin{split}
			\raisetag{1.2em}
			\texttt{b}_\texttt{1} \texttt{.f}
			\: &= \:
			\frac{1}{\theta_0} \, i^* \bigl(
				\mu_b \, \hodge \dd (\texttt{f.e}) \, \theta
				- \mu_b \, (\hodge \, \dd \, \hodge \, \upsilon) \, \texttt{s.e}
			\bigr)
			\\
			\texttt{b}_\texttt{1} \texttt{.e}
			\: &= \:
			i^*(\texttt{f.e})
		\end{split}
		\label{eq:bulk_boundary}
	\end{align}%
	\label{eq:bulk}
\end{subequations}

%
To gain insight about~\eqref{eq:bulk},
we note that
$\hodge 1 \in \Omega^n(\mathcal{Z})$
is called Riemannian volume form
since its integral yields
the volume of the manifold.
Its Lie derivative
$\mathcal{L}_{\upsilon^\sharp} (\hodge 1)$
gives the rate of local volume change.
We have
\begin{equation*}
	\mathcal{L}_{\upsilon^\sharp} (\hodge 1)
	\: = \:
	\dd( \iota_{\upsilon^\sharp} (\hodge 1) )
	\: \overset{\eqref{eq:interior_product}}{=} \:
	\dd( \hodge ( \upsilon \wedge \hodge \, \hodge 1 ) )
	\: = \:
	\dd \, \hodge \, \upsilon
\end{equation*}
Comparing to thermal conduction,
$\dd(\hodge \upsilon)$
is analogous to
$\dd(\theta)$.
Akin to how
$\kappa \, \hodge(\cdot)$
acts on
$\dd \theta = \dd(\texttt{R}_\texttt{t}\texttt{.s.e})$
to yield the heat flux,
$\mu_b \, \hodge(\cdot)$
acts on
$\dd \, \hodge \, \upsilon = \dd(\texttt{R}_\texttt{1}\texttt{.f.e})$
to yield a pressure-like quantity
counteracting volume change.
The $0$-form
$\hodge \, \dd \, \hodge \, \upsilon$
is called divergence of $\upsilon^\sharp$
and $\mu_b \geq 0$ is a bulk viscosity coefficient.
Akin to how
$\dd(\cdot)$ acts on the heat flux
to yield the net heat flux
which leads to a change of internal energy,
$\dd(\cdot)$ acts on the pressure-like quantity
to yield the net viscous force
which leads to a change of momentum.
This pattern common to
many models of relaxation processes
is quintessential for
linear irreversible thermodynamics (LIT).
It must however not lead to linear models,
since coefficients
may be functions of
quantities like temperature.
The product of
the pressure-like quantity
and the rate of volume change
$
\mu_b \,
(\hodge \, \dd \, \hodge \, \upsilon) \,
\dd \, \hodge \, \upsilon
$
is the rate at which
kinetic energy
irreversibly turns into heat.
Consequently,
the entropy production rate
is given by
$
\frac{1}{\theta} \, \mu_b \,
(\hodge \, \dd \, \hodge \, \upsilon) \,
\dd \, \hodge \, \upsilon
$.
Considering the interconnection of the subsystems for
kinetic energy,
internal energy,
and bulk viscosity,
we see that~\eqref{eq:bulk_domain}
leads to the following contributions
to the evolution equations:
\begin{align*}
	\dot{\upsilon}
	\: &= \:
	\frac{1}{\hodge m} \, \dd( \mu_b \, \hodge \, \dd \, \hodge \, \upsilon )
	\\
	\dot{s}
	\: &= \:
	\frac{1}{\theta} \, \mu_b \, (\hodge \, \dd \, \hodge \, \upsilon) \, \dd \, \hodge \, \upsilon
\end{align*}
Noting that
$
\left( \texttt{f.e}, \, \texttt{s.e} \right)
=
\left( \hodge \upsilon, \, \theta - \theta_0 \right)
$
where
$\left( \hodge \upsilon, \, \theta \right)$
is the energetic part
and
$\left( 0, \, -\theta_0 \right)$
is the entropic part,
we assert that
the energetic contribution
lies in the kernel of~\eqref{eq:bulk_domain}
and hence
energy is conserved by the irreversible process.
By invoking the integration by parts formula~\eqref{eq:partial_integration},
we obtain
\begin{equation}
	\begin{split}
		\int_\mathcal{Z} (
			&\texttt{f.e} \wedge \texttt{f.f}
			+
			\texttt{s.e} \wedge \texttt{s.f}
		)
		-
		\int_{\partial \mathcal{Z}}
			\texttt{b}_\texttt{1} \texttt{.e} \wedge
			\texttt{b}_\texttt{1} \texttt{.f}
		\: = \:
		\\
		\int_\mathcal{Z} \biggl[
			&\dd(\texttt{f.e}) \wedge \biggl(
				\frac{1}{\theta_0} \, \theta \, \mu_b \, \hodge \dd(\texttt{f.e})
			\biggr)
			\\
			-&\dd(\texttt{f.e}) \wedge \biggl(
				\frac{1}{\theta_0} \, \mu_b \, (\hodge \, \dd \, \hodge \, \upsilon) \, \texttt{s.e}
			\biggr)
			\\
			-&\texttt{s.e} \wedge \biggl(
				\frac{1}{\theta_0} \, \mu_b \, (\dd \, \hodge \, \upsilon) \, \hodge \dd(\texttt{f.e})
			\biggr)
			\\
			+&\texttt{s.e} \wedge \biggl(
				\frac{1}{\theta_0} \, \frac{1}{\theta} \, \mu_b \, (\hodge \, \dd \, \hodge \, \upsilon) \, (\dd \, \hodge \, \upsilon) \, \texttt{s.e}
			\biggr)
		\biggr]
		\: \geq \: 0
	\end{split}
	\label{eq:bulk_symmetric}
\end{equation}
which shows that~\eqref{eq:bulk}
defines a symmetric operator, etc.

\subsection{Shear Viscosity}%
\label{ssec:shear}

We understand shear viscosity
as a relaxation process
counteracting general deformation of the fluid.
Since this includes
contraction and expansion,
there is overlap with bulk viscosity.
On Euclidian spaces,
this can be avoided via
isotropic-deviatoric splitting of the stress tensor.
We carry on,
acknowledging that
shear viscosity
essentially adds more bulk viscosity.
%
While pressure and bulk viscous stress
are conveniently expressed as $0$-forms,
the non-isotropic nature of shear stress
requires a stress tensor of type
$\Omega^1(\mathcal{Z}) \otimes \Omega^{n-1}(\mathcal{Z})$,
where the first leg
represents the traction force
acting across a surface element
paired with the second leg,
see~\cite{2007KansoArroyoTongYavariMarsdenDesbrun}.
Pairing the first leg with
the velocity vector field $\upsilon^\sharp$
and integrating the remaining second leg over a surface
yields the power exchanged across the surface.
%
This motivates
the product
$
\dot{\wedge} \colon
( \Gamma(T \mathcal{Z}) \otimes \Omega^k(\mathcal{Z}) )
\times
( \Omega^1(\mathcal{Z}) \otimes \Omega^{n-k}(\mathcal{Z}) )
\rightarrow \Omega^n(\mathcal{Z})
$
which pairs the first legs
and uses $\wedge$ on the second legs.
We implicitly use
the isomorphism
$\Gamma(T \mathcal{Z}) \simeq \Gamma(T \mathcal{Z}) \otimes \Omega^0(\mathcal{Z})$.
%
Analogous to how
$\dd(\hodge \, \upsilon) \in \Omega^n(\mathcal{Z})$
is the rate of volume change,
$
\nabla( \upsilon^\sharp )
\in \Gamma(T \mathcal{Z}) \otimes \Omega^1(\mathcal{Z})
$
is the strain rate.
The covariant derivative $\nabla$
measures how
$\upsilon^\sharp \in \Gamma(T \mathcal{Z})$
changes
in the direction given by a vector
paired with the second leg.
%
Similar to how
$\mu_b \, \hodge ( \dd(\hodge \, \upsilon) ) \in \Omega^0(\mathcal{Z})$
is the bulk viscous stress,
$
T = \mu_s \, \hodge_2(\mathrm{sym}(\flat_1( \nabla( \upsilon^\sharp ))))
\in \Omega^1(\mathcal{Z}) \otimes \Omega^{n-1}(\mathcal{Z})
$
is the shear stress.
Subscripts
$1$ and $2$
indicate the leg on which $\flat$ and $\hodge$ act
and $\mathrm{sym}$ symmetrizes the two legs.
%
Akin to how
$\dd( \mu_b \, \hodge ( \dd(\hodge \, \upsilon) ) ) \in \Omega^1(\mathcal{Z})$
is the net bulk viscous stress,
$
\dd_\nabla(T)
\in \Omega^1(\mathcal{Z}) \otimes \Omega^n(\mathcal{Z})
$
is the net shear stress
acting on a volume element paired with the second leg.
%
The covariant exterior derivative $\dd_\nabla$
is the formal adjoint of $-\nabla$
as shown by the integration by parts formula
\begin{equation}
	\begin{split}
		\raisetag{4.3em}
		\int_\mathcal{Z}
		\dd ( \upsilon^\sharp \dot{\wedge} T )
		\: &= \:
		\int_\mathcal{Z}
		\upsilon^\sharp
		\dot{\wedge}
		\dd_\nabla T
		\: + \:
		\int_\mathcal{Z}
		\nabla ( \upsilon^\sharp )
		\dot{\wedge}
		T
		\\
		\: &= \:
		\int_{\partial \mathcal{Z}}
		i^*_2(\upsilon^\sharp \dot{\wedge} T))
		\: = \:
		\int_{\partial \mathcal{Z}}
		i^*_2(\upsilon^\sharp) \dot{\wedge} i^*_2(T))
		\,,
	\end{split}%
	\label{eq:exterior_covarint_derivative}
\end{equation}
where the boundary terms are two-point tensors
since only the second legs can be pulled back to $\partial \mathcal{Z}$.
%
The composition pattern
is given by
the following expression:
\begin{center}
	\includegraphics[width=7.3cm]{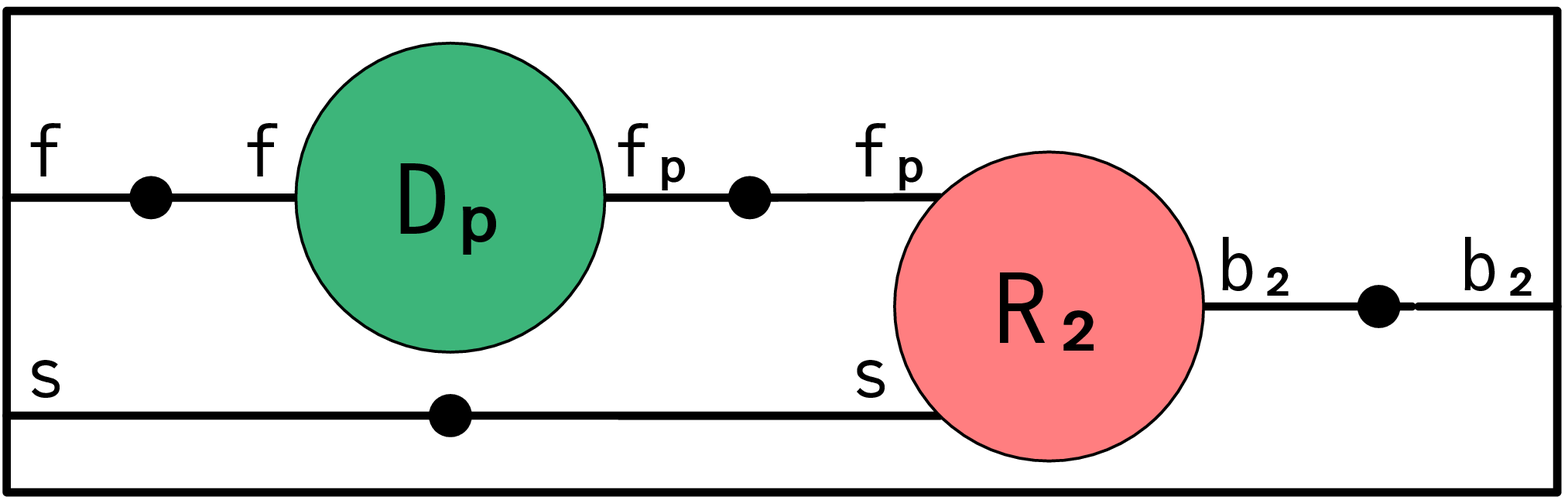}
\end{center}
Port \texttt{f} is related to
the viscous force,
\texttt{s} is related to entropy production,
and $\texttt{b}_\texttt{2}$ is related to
viscous stress acting across $\partial \mathcal{Z}$.
While
$\texttt{R}_\texttt{2}$
represents the resistive structure
defining the irreversible process,
$\texttt{D}_\texttt{p}$
represents the Dirac structure
which translates between
stress $T$
at port $\texttt{f}_\texttt{p}$
and the resulting force
$f \in \Omega^1(\mathcal{Z})$
at port \texttt{f}.
%
Since we have
\begin{equation*}
	\int_\mathcal{Z}
	\hodge \upsilon \wedge f
	=
	\int_\mathcal{Z}
	\sigma \, (\iota_{\upsilon^\sharp} f) \, (\hodge 1)
	=
	\int_\mathcal{Z}
	\upsilon^\sharp \dot{\wedge} f \otimes (\sigma \, \hodge 1)
	=
	\int_\mathcal{Z}
	\upsilon^\sharp \dot{\wedge} T
	\,,
\end{equation*}
we can conclude that
the following relation
defines
the Dirac structure
associated to $\texttt{D}_\texttt{p}$:
\begin{equation}
	\left[
		\begin{array}{c}
			\texttt{f.f} \\
			\hline
			\texttt{f}_\texttt{p} \texttt{.e} \\
		\end{array}
	\right]
	\: = \:
	\left[
		\begin{array}{c|c}
			0 & \sigma \, \hodge_2(\cdot) \\
			\hline
			\sigma \, \sharp( \hodge( \cdot)) & 0
		\end{array}
	\right]
	\,
	\left[
		\begin{array}{c}
			\texttt{f.e} \\
			\hline
			\texttt{f}_\texttt{p} \texttt{.f}
		\end{array}
	\right]%
	\label{eq:shear_dirac_force_stress}
\end{equation}
%
The resistive structure associated to
$\texttt{R}_\texttt{2}$
is defined by the following relation:
\begin{subequations}
	\begin{align}
		\begin{split}
			\raisetag{2em}
			\left[
				\begin{array}{c}
					\texttt{f}_\texttt{p} \texttt{.f} \\
					\texttt{s.f}
				\end{array}
			\right]
			&=
			\frac{1}{\theta_0}
			\left[
				\begin{array}{cc}
					A & B \\
					C & D
				\end{array}
			\right]
			\left[
				\begin{array}{c}
					\texttt{f}_\texttt{p} \texttt{.e} \\
					\texttt{s.e}
				\end{array}
			\right]
			\\
			A &= -\dd_\nabla \bigl( \mu_s \, \hodge_2(\mathrm{sym}(\flat_1(\nabla(\cdot)))) \, \theta \bigr) \\
			B &=  \dd_\nabla \bigl( \mu_s \, \hodge_2(\mathrm{sym}(\flat_1(\nabla(\upsilon^\sharp)))) \, (\cdot) \bigr) \\
			C &= -\mu_s \, \nabla(\cdot) \dot{\wedge} \hodge_2(\mathrm{sym}(\flat_1(\nabla(\upsilon^\sharp)))) \\
			D &= \frac{1}{\theta} \, \mu_s \nabla(\upsilon^\sharp) \dot{\wedge} \hodge_2(\mathrm{sym}(\flat_1(\nabla(\upsilon^\sharp)))) \, (\cdot)
		\end{split}
		\label{eq:shear_domain}
		\\
		\begin{split}
			\raisetag{1.2em}
			\texttt{b}_\texttt{2} \texttt{.f}
			\: &= \:
			\frac{1}{\theta_0} \, i^* \bigl(
				\mu_s \, \flat_1(\hodge_2(\mathrm{sym}(\nabla(\texttt{f}_\texttt{p} \texttt{.e})))) \, \theta \\
			   & \quad \quad \quad
			   - \mu_s \, \flat_1(\hodge_2(\mathrm{sym}(\nabla(\upsilon^\sharp)))) \, (\texttt{s.e})
			\bigr)
			\\
			\texttt{b}_\texttt{2} \texttt{.e}
			\: &= \:
			i^*(\texttt{f}_\texttt{p} \texttt{.e})
		\end{split}
		\label{eq:shear_boundary}
	\end{align}%
	\label{eq:shear}
\end{subequations}

\subsection{Composed Model}%
\label{ssec:composed}

The interconnection of all five subsystems yields
the following system of equations
which must be satisfied on $\mathcal{Z}$:
\begin{subequations}
	\begin{align}
		\begin{split}
			\dot{\upsilon}
			\: &= \:
			-\iota_{\upsilon^\sharp}(\dd \upsilon)
			-\dd (\frac{1}{2} \hodge (\upsilon \wedge \hodge \upsilon)) \\
			& \quad \quad + \frac{1}{\hodge m} \, \dd( \mu_b \, \hodge \, \dd \, \hodge \, \upsilon ) \\
			& \quad \quad + \frac{1}{\hodge m} \, \hodge_2 \, \dd_\nabla \bigl( \mu_s \, \hodge_2(\mathrm{sym}(\nabla(\upsilon))) \bigr) \\
			\label{eq:composed_momentum_balance}
		\end{split}
		\\
		\dot{m}
		\: &= \:
		-\dd (\hodge m \, \hodge \upsilon)
		\label{eq:composed_mass_balance}
		\\
		\begin{split}
			\dot{s}
			\: &= \: -\dd ( \hodge s \, \hodge \upsilon ) \\
			& \quad \quad + \frac{1}{\theta} \, \dd( \kappa \, \hodge \dd \theta ) \\
			& \quad \quad + \frac{1}{\theta} \, (\hodge \, \dd \, \hodge \, \upsilon) \, \mu_b \, \dd \, \hodge \, \upsilon \\
			& \quad \quad + \frac{1}{\theta} \, \nabla(\upsilon^\sharp) \dot{\wedge} \bigl( \mu_s \, \hodge_2(\mathrm{sym}(\nabla(\upsilon))) \bigr)
			\label{eq:composed_entropy_balance}
		\end{split}%
	\end{align}%
	\label{eq:composed}%
\end{subequations}
The boundary ports labeled $\texttt{b}_i$
with $i \in \{ \texttt{k}, \, \texttt{s}, \, \texttt{m}, \, \texttt{t}, \, \texttt{1}, \, \texttt{2} \}$
defined in the previous subsections
imply the boundary conditions on $\partial \mathcal{Z}$.

\section{Conclusion}%
\label{sec:conclusion}

Using the Exergetic Port-Hamiltonian Systems (EPHS) framework,
we present a structured model of
the Navier-Stokes-Fourier fluid.
The model has
five subsystems for
kinetic energy storage,
internal energy storage,
thermal conduction,
bulk viscosity,
and shear viscosity.
The composed model comprises
nine primitive components,
namely
two for exergy storage,
four representing Dirac structures,
and three representing resistive structures.
The power-preserving interconnection of
subsystems, components, and boundary ports
is defined explicitly
by composition patterns
stated in the EPHS syntax.
The first and the second law of thermodynamics
is implied by
the EPHS semantics.
The distributed-parameter model
is expressed
using Cartan's exterior calculus
on Riemannian manifolds.
%
Future work should aim to
define model discretization
as a natural transformation between functors
assigning distributed-parameter and lumped-parameter semantics,
since this would imply
preservation of the compositional and thermodynamic structure.
Once this is achieved,
not much would be missing
before one could apply the model
as a building block in larger systems.



\bibliography{literature}


\end{document}